\documentclass[letterpaper]{mc2023}
\usepackage{tabls}
\usepackage{cites}
\usepackage{epsf}
\usepackage{appendix}
\usepackage{ragged2e}
\usepackage{siunitx}
\usepackage{algorithm}
\usepackage[top=1in, bottom=1in, left=1in, right=1in]{geometry}
\usepackage{enumitem}
\setlist[itemize]{leftmargin=*}
\usepackage{caption}
\title{Hybrid-Delta Tracking on a Structured Mesh in MCATK \\
  }
\author{%
  \textbf{J. P. Morgan$^{1,2, \dagger}$, Travis J. Trahan$^{2, *}$, Timothy P. Burke$^2$,}\\ 
  \textbf{Colin J. Josey$^2$, and Kyle E. Niemeyer$^{1}$} \vspace{6pt} \\
  $^1$School of Mechanical Industrial and Manufacturing Engineering
     \\ Oregon State University,
     204 Rogers Hall, Corvallis, OR 97331\vspace{6pt} \\ 
  $^2$XCP-3\\ 
  Los Alamos National Laboratory, P.O. Box 1663, Los Alamos, NM 87545\vspace{6pt} \\
  $^\dagger$\url{morgjack@oregonstate.edu}, $^*$\url{tjtrahan@lanl.gov}
}
\newcommand{\authorHead}{Morgan, J. P. et al.}
\newcommand{\shortTitle}{Hybrid-Delta Tracking}
\begin{document}
\maketitle
\justify 
\parskip 6pt plus 1 pt minus 1 pt

\begin{abstract}    
Monte Carlo Application Toolkit (MCATK) commonly uses surface tracking on a structured mesh to compute scalar fluxes. In this mode, higher fidelity requires more mesh cells and isotopes and thus more computational overhead --- since every time a particle changes cells, new cross-sections must be found for all materials in a given cell --- even if no collision occurs in that cell. We implement a hybrid version of Woodcock (delta) tracking on this imposed mesh to alleviate the number of cross-section lookups. This algorithm computes an energy-dependent microscopic majorant cross section is computed for the problem. Each time a particle enters a new cell, rather than computing a true macroscopic cross-section over all isotopes in the cell, the microscopic majorant cross-section is simply multiplied by the total number density of the cell to obtain a macroscopic majorant cross-section for the cell. Delta tracking is then performed within that single cell. This increases performance with minimal code changes, speeding up the solve time by a factor of 1.5---1.75 for k-eigenvalue simulations and 1.2---1.6 for fixed source simulations in a series of materially complex criticality benchmarks.
\end{abstract}
\vspace{6pt}

\section{INTRODUCTION} 
\label{sec:intro}

The Monte Carlo Application ToolKit (MCATK) \cite{MCATK} is a Monte Carlo particle transport code that commonly uses surface tracking on a structured mesh (e.g., a Cartesian grid) to compute scalar fluxes.
The structured mesh makes distance-to-boundary calculations cheap compared to simulations using a constructive solid geometry.

Many current simulations need a great number of isotopes to be modeled properly, such that a single mesh cell might contain many isotopes. 
MCATK's standard algorithm computes the total cross-section of all isotopes in a cell every time a particle moves between cells.
Computationally expensive lookup and interpolation functions are called at every boundary crossing.
When cell sizes are small compared to a neutron mean free path, often this cross section lookup finds that a particle does not collide in the cell.
This process repeats until a collision occurs.

In this work, we introduce a hybrid-delta tracking algorithm to eliminate the macroscopic cross section lookup at surface crossings while still performing standard surface tracking.

\section{HYBRID-DELTA TRACKING}
\label{sec:method}

Woodcock, or delta, tracking \cite{woodcock1965} is a variance-reduction technique that computes the majorant cross-section for the whole problem space, then uses this to determine a distance to collision for all particles.
Coupled with rejection sampling to sort for phantom collisions, and a collision estimator to compute scalar flux, delta tracking often improves performance over analogue Monte Carlo in problems that warrant it.
Many production Monte Carlo Neutron transport codes like Serpent \cite{Serpent2013} use this method.

We implement a hybrid surface-delta tracking algorithm to reduce the number of cross-section lookups while still using MCATK's surface tracking on a structured mesh to find scalar flux. First we start by computing the microscopic majorant cross-section over the whole problem space, like in traditional delta tracking:
\begin{equation}
    \label{eq:majorant}
    \sigma_{M}(E) = \max\left(\sigma_{T,k}(E), ..., \sigma_{T,K}(E)\right) \,\text{,}
\end{equation}
where $E$ is energy, $\sigma_{M}$ is the microscopic majorant cross-section, and $\sigma_{T,k}$ is the microscopic total cross-section of the $k^{\text{th}}$ material. In this work, we use one majorant over the whole problem regardless of location, but regional microscopic majorants could still be used in future work. Now to sample a distance we calculate the distance to a collision
\begin{equation}
    \label{eq:sample}
    D = \frac{-\ln{\xi}}{N_i \sigma_{M}(E)} \, \text{,} 
\end{equation}
where $\xi$ is a random number between zero and one and $N_i$ is the number density in cell $i$.
Thus far the number density of the cell is the only parameter that depends on the particle's location in the problem, so there is no need to look up and interpolate total cross-sections when tallying the distance traveled through the mesh before a collision.
This is equivalent to assuming all atoms in the cell are of the type with the largest microscopic total cross-section at that energy.

If the potential collision occurs within the cell, we move the particle to the sampled distance and do rejection sampling, since we are now potentially forcing collisions that did not occur. We sort out these phantom collisions by allowing particles to continue to a new sampled distance if
\begin{equation}
    \label{eq:reject}
    \xi < \frac{ \Sigma_{T,i}(E) } { N_i \sigma_M(E) } \, \text{,}
\end{equation}
where $\xi$ is a new random number between zero and one and $\Sigma_{T,i}(E)$ is the macroscopic total cross-section of the cell. 
%not sure about these indices but I want to denote that this is in a new cell and not necessarily the one next door

Since we are still moving the particles through each cell, we can use a track-length tally for estimating scalar flux.
This may not offer as much raw speed-up as full delta tracking, but may result in greater efficiency, as a track-length estimator has a lower variance than a collision estimator \cite{mc2018}.

\begin{figure}[!htb]
  \centering
  \includegraphics[scale=0.95]{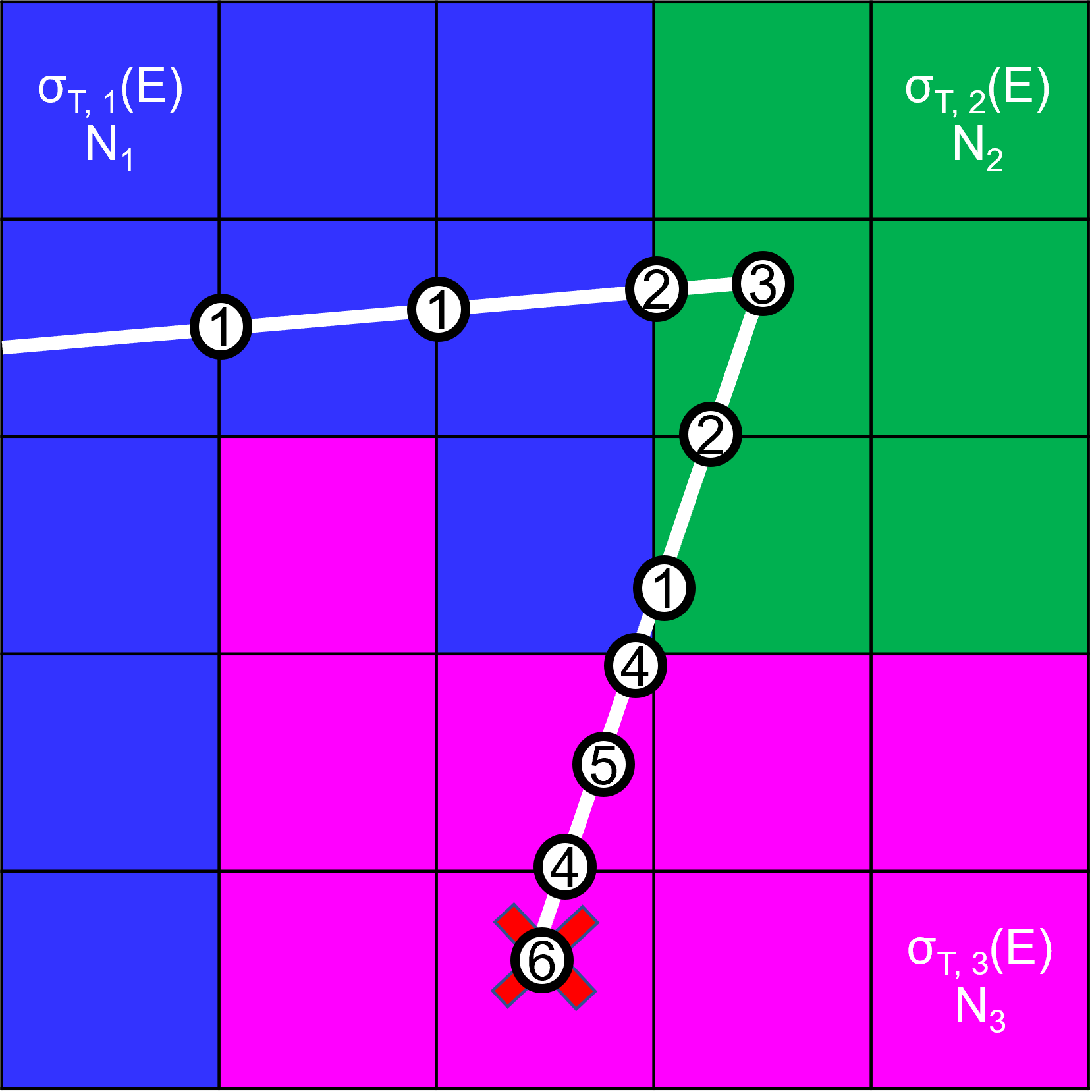}
  \caption{A material region in a structured mesh where each color represents a different material with a different $\boldsymbol{N_i}$ and $\boldsymbol{\sigma_T(E)}$}
  \label{fig:algo}
\end{figure}

Consider the region with three materials shown in blue, green, and pink presented in Figure \ref{fig:algo}. A particle enters the region from the left and undergoes transport with operations required at every distance to a boundary or collision:
\begin{enumerate}
    \item the particle crosses into region 1 requiring that $\sigma_M(E)$ (pre-computed) is interpolated, multiplied by $N_{1}$ to form $\Sigma_{M,1}$, and a new distance to collision is computed;
    \item the particle crosses into region 2 requiring that $\sigma_M$ (pre-computed) is interpolated then multiplied by $N_{2}$ to form $\Sigma_{M,2}$, and a new distance to collision is computed;
    \item the distance to collision is small enough to place a collision in the cell. Rejection sampling is done and it is found that this was a real event. Particle undergoes a scattering event;
    \item the particle crosses into region 3 requiring that $\sigma_M$ (pre-computed) is interpolated then multiplied by $N_{3}$ to form $\Sigma_{M,3}$, and a new distance to collision is computed;
    \item the distance to collision  is small enough to place a collision in the cell. A potential collision is rejected; and
    \item the distance to collision  is small enough to place a collision in the cell. Rejection sampling determines that the collision is real, and the particle is absorbed.
\end{enumerate}
Here, $\sigma_M = \max{(\sigma_{T,1}, \sigma_{T,2}, \sigma_{T,3})}$. $\sigma_{T,i}$ is only looked up for rejection sampling, required in operations 3, 5 and 6. For this hypothetical problem we have reduced the number of lookups to find $\sigma_{T,i}$ from ten to three. Depending on the number of isotopes in the cell, this could represent a significant number of operations.
Speed-up in MCATK with this method comes from the avoidance of these lookup functions to find the dominant total microscopic cross section between \textit{all} the isotopes that are contained within a cell. 
We expect a significant speed-up as this function is called frequently in standard surface tracking. 
We also expect that the speedup will be greater when there are more isotopes within a cell. 
All other operations remain the same between standard tracking and hybrid surface-delta tracking on the structured mesh.

Extending our implementation of the hybrid-delta tracking scheme in models that use constructive solid geometry instead of structured meshes will require no additional work.
Since small cell sizes in structured meshes is analogous to complex and small geometries in constructive solid geometry, we expect similar performance increases when these models also have complex isotopic compositions.

We first implemented hybrid-delta tracking in MCATK's k-eigenvalue algorithm then implemented it for fixed source problems.
Since all of MCATK's algorithms (e.g., k-eigenvalue, fixed source, etc.) use the same collision kernel no alterations were required to move from an implementation in one algorithm to the next other than Boolean switches to initialize the mini-delta tracking algorithm.

\section{BENCHMARKS}
\label{sec:benchmarks}
We investigated the performance of hybrid-delta tracking on a series of benchmark problems: one from the Godiva IV experiment (HEU-MET-FAST-086, case 5) and two from the MUSiC experiment.
All three are materially complex, having between 48 and 125 isotopes over the whole model, and implement a structured tracking mesh imposed on the geometry. 
The minimum mesh cell dimension is 1 mm in the fissile region of each problem.

\begin{figure}[!htb]
  \centering
  \includegraphics[scale=0.6]{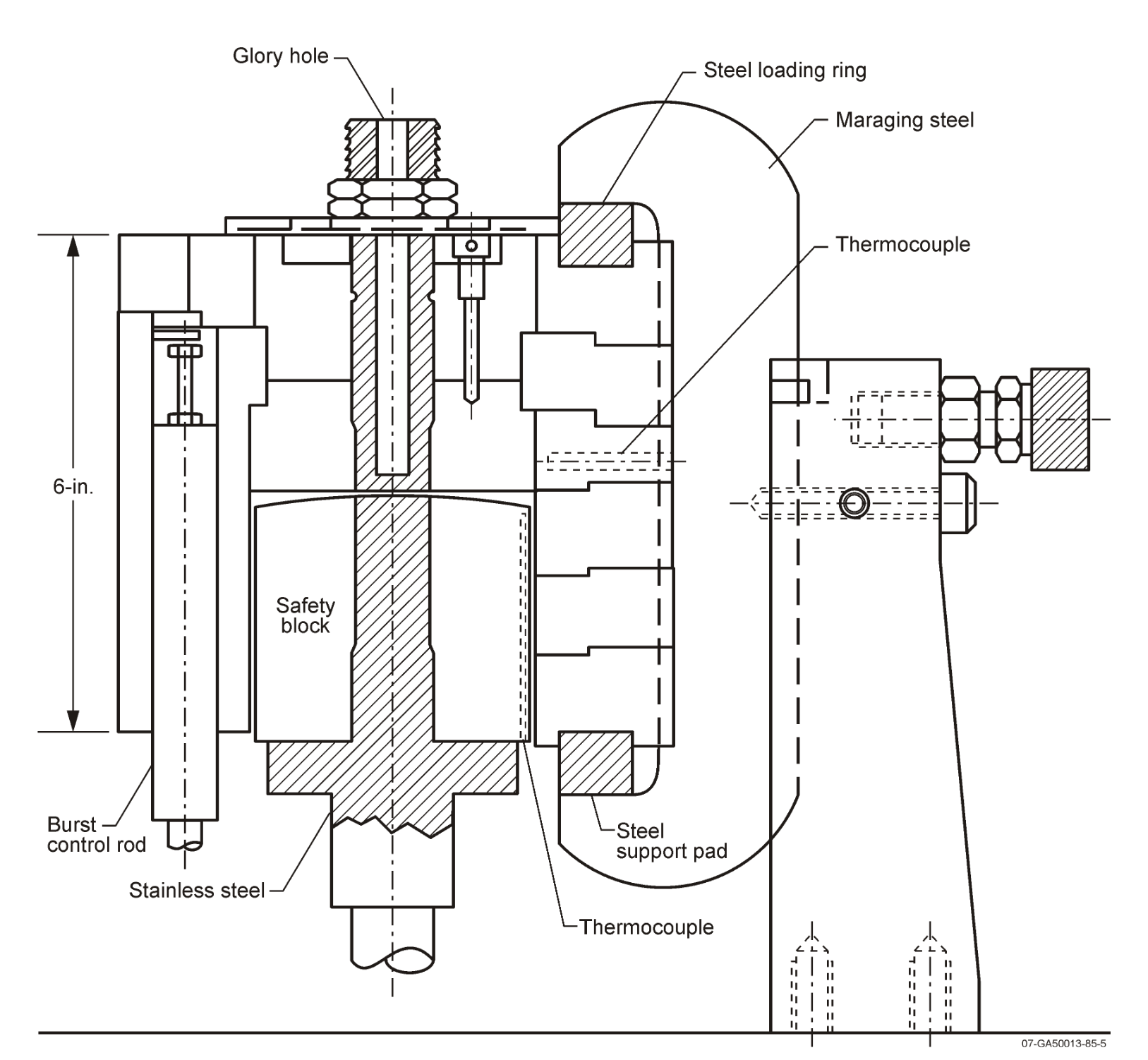}
  \caption{Cutaway view of the Godiva IV core and its restraints used in the HEU-MET-FAST-086 benchmark \cite{godiva2014}. }
  \label{fig:godiva}
\end{figure}

%\subsection{Godiva IV}
%\label{subsec:godiva}
The Godiva IV super-prompt-critical burst experiments use two control rods and one burst rod made of highly enriched uranium and molybdenum to control criticality \cite{osti_9564352009}.
As the rods are inserted reactivity goes up, and vice versa when they are removed.
The only fission source is the spontaneous fission from the highly enriched uranium in the experiment.
Figure \ref{fig:godiva} shows a cross section of the Godiva IV core assembly and restraints which are modeled in our benchmark.
This is the simplest benchmark we model, requiring the fewest number of isotopes. 

%\subsection{MUSiC}
%\label{subsec:music}

\begin{figure}[!htb]
  \centering
  \includegraphics[scale=0.9]{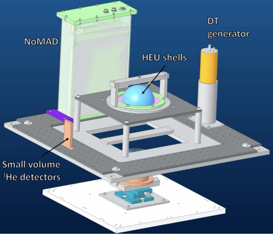}
  \includegraphics[scale=0.9]{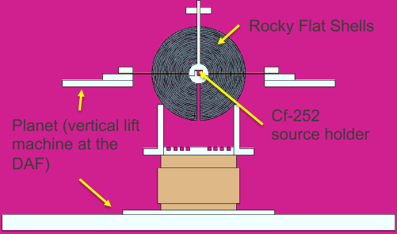}
  \caption{Left: Overall configuration of the MUSiC experiment. Right: Schematic of the Rocky Flats highly enriched uranium hemispherical shells \cite{osti_1781360} }
  \label{fig:music}
\end{figure}

The Measurement of Uranium Subcritical and Critical (MUSiC) experiments (IER 488) use stack-able hemispheres of highly enriched uranium --- known as the Rocky Flats Shells --- to take data with various detectors \cite{music2021}.
Figure \ref{fig:music} at right shows the highly enriched uranium shells which are about \SI{0.3}{\centi\meter} thick and have variable radii between about \SI{2}{\centi\meter} and \SI{10}{\centi\meter}.
Figure \ref{fig:music} at left shows the full configuration of the experiment with fission sources and detectors.
Fission is induced with a Cf-252 source at the center of the shell array as well as an external deuterium source.

\section{PERFORMANCE RESULTS} 
\label{sec:results}

We verified all models on both k-eigenvalue and fixed source problems with the standard tracking algorithm from MCATK, then computed the speed-up from our hybrid-delta tracking scheme.
To produce timing results these benchmarks ran with 1 rank; however, parallelization is already enabled for hybrid-delta tracking in MCATK via MPI.
Material, geometry, and mesh between k-eigenvalue and fixed source simulations are the same for a given case.

\subsection{k-eigenvalue Simulation}
K-eigenvalue simulations were started with 100 inactive cycles before 500 active ones using \num{1e5} particles in each cycle.
Table \ref{table:runtime} shows the performance increase when hybrid-delta tracking is enabled.
The differences in eigenvalue between simulations with and without the hybrid-delta tracking algorithm are all within 1.12 standard deviations. This table also shows a significant speed-up in the over all solve time of MCATK with delta tracking incurring between a 1.54 --- 1.75$\times$ speed-up.

\begin{table}[!htb]
  \centering
  \caption{\bf Benchmark results: where $\boldsymbol{\Delta \sigma}$ is the difference in number of standard deviations of $\boldsymbol{k_{\text{eff}}}$ between the standard algorithm and hybrid-delta tracking.}
  \label{table:runtime} 
  \begin{tabular}{c c c c c c  } \hline 
    Test Model & MCATK & MCATK & $k$ & $\Delta \sigma$ & speed-up\\
               & Surface (s)  & Hybrid-Delta (s) &  & &\\ \hline
    \ Godiva Case 5 & 16900 &  10987 & 0.99736 &  -0.747 & 1.54 \\
    \ MUSiC Case 8  & 23937 &  14832 & 0.99970 &  0.130 & 1.61 \\
    \ MUSiC Case 9  & 22649 &  12973 & 0.99929 &  -1.105 & 1.75 \\ 
    \hline
  \end{tabular}
\end{table}

\subsection{Fixed Source Simulations}

To compute error between tracking schemes in fixed source simulations we used the estimates of the $\alpha$-eigenvalue computed using MCATK's time-dependent algorithm.
The benchmarks were started at \SI{0}{\second} and ran to \SI{500e-8}{\second} with a time step of $\Delta t =$ \SI{1e-8}{\second}.
The particle population was combed between every time step up or down to \num{1e5} particles. 
Table \ref{table:runtime_trans} shows less speed-up then for k-eigenvalue computations (only between 1.24$\times$ and 1.63$\times$) but still significant for minimal alterations to a production code. This table also shows the average $\alpha$-eigenvalues method within three standard deviations.

\begin{table}[!htb]
  \centering
  \caption{\bf Benchmark results: where $\boldsymbol{\Delta \sigma}$ is the difference in number of standard deviations of the $\alpha$-eigenvalues between standard algorithm and hybrid-delta tracking.}
  \label{table:runtime_trans} 
  \begin{tabular}{c c c c c c } \hline 
    Benchmark & MCATK            & MCATK         & $\alpha_{\text{avg}}$ & $\Delta \sigma$ & speed-up\\
               & Surface (s)  & Hybrid-Delta (s) & &                 &\\ \hline
    \ Godiva Case 5 & 2689 &  2168 & \num{-3.54e-3} &  \num{-1.91} & 1.24 \\
    \ MUSiC Case 8  & 4352 &  2665 & \num{-1.26e-3} &  \num{-0.76} & 1.63 \\
    \ MUSiC Case 9  & 4440 &  2786 & \num{-1.02e-3} &  \num{-2.75} & 1.59 \\ 
    \hline
  \end{tabular}
\end{table}

\begin{figure}[!p]
  \centering
  \includegraphics[scale=.92]{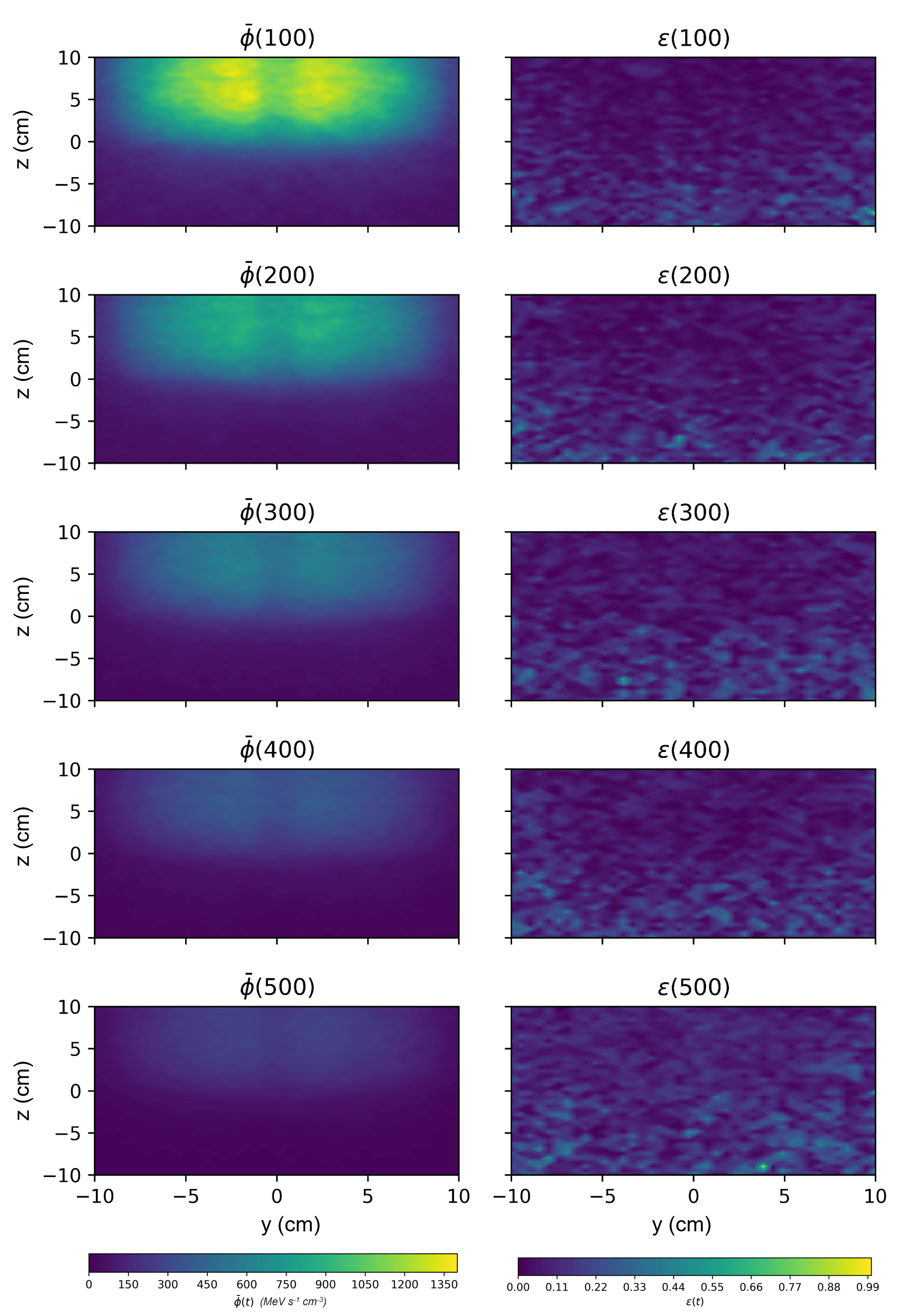}
  \caption{Right: Spatial averaged scalar flux of Godiva IV (HEU-MET-FAST-086) from MCATK. Left: Relative error between solution from standard tracking algorithm and hybrid-delta tracking. Shown through time (every $\boldsymbol{100 * 10^{-8}}$s).}
  \label{fig:godiva_results}
\end{figure}

\DeclareSIUnit\electronvolt{e\kern-.15em V}

Figure \ref{fig:godiva_results} at left shows spatial averaged scalar flux ($\bar{\phi}$) of the \num{0.1} to \SI{1.0}{\mega\electronvolt} energy bin on an $y-z$ plane slice at $x=$ \SI{0}{\centi\meter} in a high fidelity version (same simulation but with \num{1e8} particles per time step) of Godiva IV as a fixed source problem. The source in this simulation is an isotropic burst of neutrons at $(0,0,0)$ \SI{}{\centi\meter} and $t=0.0$ \SI{}{\second} with an energy of \SI{1}{\mega\electronvolt}.
At right Figure \ref{fig:godiva_results} shows the relative error
\begin{equation}
    \epsilon = \frac{\left|\bar{\phi}_{\Delta} - \bar{\phi}_{\text{Standard}}\right|}{\bar{\phi}_{\Delta}},
\end{equation}
where $\bar{\phi}_{\Delta}$ is the spatial averaged scalar flux solution from the hybrid-delta tracking scheme and $\bar{\phi}_{\text{Standard}}$ is from the standard tracking scheme.
It shows that the relative error in this slice is small between the two methods of tracking.
This further demonstrates that a hybrid-delta tracking scheme is not biasing the results of MCATK's standard surface tracking algorithm in these models.

\section{CONCLUSIONS \& FUTURE WORK}
Hybrid-delta tracking on a structured mesh improves run time in large and materially complex simulations like the ones we benchmarked: 1.5---1.75$\times$ speed-up for k-eigenvalue and 1.2---1.6$\times$ speed-up for fixed source problems.
We expect further optimizations will improve speed-up for these algorithms.

The solutions found with hybrid-delta tracking match to within three standard deviations of solutions found with MCATK's standard algorithm. 
The advantages of this algorithm over full delta tracking on structured meshes are that track-length estimators may still be used, and that minimal the changes are required to the code as the treatment of the geometry does not change.

Further work is required to verify and validate the hybrid-delta tracking scheme.
For k-eigenvalue problems we will use the criticality validation suite for MCNP \cite{mcnpCriticality} as well as other criticality benchmarks such as the Sub-critical Copper-Reflected $\alpha$-phase Pu (SCR$\alpha$P) experiment \cite{ICSBEP}.
For fixed source solutions we will verify with the analytical AZURV1 \cite{Ganapol2001HomogeneousBenchmarks} benchmark problems. We also plan to study how the algorithm performs as a function of time step size, since small time steps can suffer from poor performance in a similar manner as small cell sizes.

We expect that this work will increase the overall performance of MCATK when delta tracking is appropriately used in problems that warrant it.

\section*{ACKNOWLEDGEMENTS}
Research presented in this article was supported by the Laboratory Directed Research and Development program of Los Alamos National Laboratory under project number 20220084DR.

This work was supported by the Center for Exascale Monte-Carlo Neutron Transport (CEMeNT) a PSAAP-III project funded by the Department of Energy, grant number: DE-NA003967.

The authors would like to R. Arthur Forster for the discussions that inspired this work as well as Theresa Cutler, Travis Smith, and Robert Weldon Jr. for providing benchmark problem input files.

LA-UR-23-24971

\setlength{\baselineskip}{12pt}
\bibliographystyle{mc2023}
\bibliography{mc2023}

\end{document}